# Digital Therapeutics for Mental Health:

# Is Attrition the Achilles Heel?


Adaora Nwosu[1], Samantha Boardman[2], Mustafa M. Husain[3], P. Murali Doraiswamy[1, 4]

1. Department of Psychiatry and Behavioral Sciences, Duke University School of Medicine

2. Department of Psychiatry, Weill Cornell Medical College

3. Departments of Psychiatry, Neurology and Biomedical Engineering, The University of Texas, Southwestern Medical Center

4. Duke Institute for Brain Sciences, Duke University




**ABSTRACT**

Digit therapeutics are novel software devices that clinicians may utilize in delivering quality mental health care and ensuring positive outcomes. However, uptake of digital therapeutics and clinically tested software-based programs remains low. This article presents possible reasons for attrition and low engagement in clinical studies investigating digital therapeutics, analyses of studies in which engagement was high, and design constructs that may encourage user engagement. The aim is to shed light on the importance of real-world attrition data of digital therapeutics, and important characteristics of medical devices that have positively influenced user engagement. The findings presented in this article will be useful to relevant stakeholders and medical device experts tasked with addressing the gap between software medical design and user engagement present in digital therapeutic clinical trials.

**INTRODUCTION**

Large unmet needs in mental health combined with the stress caused by pandemic mitigation measures have accelerated the use of digital mental health apps and software-based solutions (1). Global investor funding for virtual behavioral services and mental health apps in 2021 exceeded $5.5 billion, a 139% jump from 2020, according to CB Insights (2). While there are thousands of apps claiming to improve various aspects of mental wellbeing, many of them have never gone through clinical trials or regulatory scrutiny. The term "digital therapeutic" is used in the literature to distinguish high quality evidence-based software programs from wellness apps (3). Regulators use the term "software as a medical device" (SaMD) or "software in a medical device" (SiMD) to refer to software that functions as a medical device and is promoted to treat a specific condition. When a SaMD or SiMD is deployed on a phone it is referred to as a mobile



medical app (MMA) (4,5). The International Medical Device Regulators Forum (IMDRF), a voluntary group of medical device regulators from around the world has developed detailed guidance on definitions, framework for risk categorization, quality management, and the clinical evaluation of such devices (6-8). Non-traditional approaches, outside of RCTs, to evaluate efficacy for such tools has also been discussed elsewhere (9).

To date, only four clinically tested software devices have been authorized by the U.S. Food and Drug Administration for treating specific mental health disorders (excluding devices marketed under pandemic-related emergency use authorization). These include reSET for substance abuse disorder (10), reSET-O for opioid use disorder (11), Somryst for chronic insomnia (12,13) and EndeavorRx for pediatric attention deficit hyperactivity disorder (14,15). SaMDs and MMAs for treating mild cognitive impairment, Alzheimer's disease, schizophrenia, autism, depression, social anxiety disorder, phobias and PTSD are in clinical trials (1, 3, 5, 17) and may also come to market soon. The state of efficacy for non-regulated, wellness apps (e.g. for mindfulness or stress management) is beyond the scope of this article, and readers are referred elsewhere for the state of efficacy with these apps (16).

## HIGH ATTRITION AND LOW ENGAGEMENT

While digital therapeutics and apps undoubtedly hold promise, relatively little attention has focused on attrition rates. Even effective apps will have limited impact if they are not highly engaging and result in high attrition (18, 19). Attrition, the loss of a randomized subject(s) from a study sample, is a very common issue in clinical trials and results from several causes such as refusal to participate after randomization, an early dropout from the study, and loss of subject's



study data.  Attrition can substantially bias estimates of efficacy and reduce generalizability (20).

Traditionally, regulatory trials of psychopharmacological agents have used the last observation

carried forward (LOCF) statistical method to accommodate attrition – but this has been

increasingly replaced with mixed-effects models, and pattern-mixture and propensity

adjustments (20).  Compliance in trials of psychopharmacological agents is traditionally

measured via pill counts.  However, in virtual platform trials of digital therapeutics, compliance

cannot simply be measured by the number of times a subject logs on to an app and it is important

to also measure and report how engaged users were with the app (21).  Currently, there is no

standard way to define what constitutes meaningful engagement and how to compare

engagement across different digital therapeutic devices (21).  There is also no consensus as to

how to deal with users who are non-engaged but stay in the study.

As patients typically use apps on their own time, they must be intrinsically motivated to do so

and must perceive the benefits from the app as meaningful (18). Such intrinsic motivation may

be low for psychiatric patients with depression, anhedonia, or cognitive difficulties.  For

example, in one study of internet-based cognitive behavioral therapy for depression, the highest

engagers comprised just 10.6% of the sample (22).  This is further highlighted by a 2020 meta-

analysis of 18 randomized trials of (non-FDA cleared) mobile apps for treating depression (trial

duration ranging from 10 days to 6 months), in which the pooled dropout rate was 26.2% and

rose to 47.8% when adjusted for publication bias (19).  The authors concluded that this raises

concern over whether efficacy was overstated in these studies.  Real-world attrition rates for non-

FDA cleared mental wellness apps are not readily available for direct comparison. But one study

of 93 non-FDA approved Android apps (median installs 100,000), targeting mental wellbeing,



found the medians of app 15-day and 30-day retention rates were very low at 3.9% and 3.3% (23). The median percentage of daily active users (open rate) was only 4.0%. These data highlight that the number of app installs has very little correlation with daily long-term usage.

Attrition and engagement rates (self-defined by study sponsors) in the pivotal studies for the four FDA-authorized neuropsychiatric digital therapeutics are shown in Table 1. The studies reported significant benefits for the digital therapeutic versus a control condition (Table 1, 10-15). Sample sizes were adequate, ranging from 170 to 1149 participants (Table 1). Active intervention durations were relatively short ranging from four to 12-weeks (Table 1). Trial design, nature of therapy, incentives, and diagnosis influenced attrition. The Somyrst trials additionally reported 6-month and 12-month follow up data. Attrition was lowest and compliance was highest in the pivotal study (14) of EndeavorRx for pediatric ADHD (Table 1) – this was a short 4-week trial of an interactive videogame where compliance was monitored electronically and there was close parental supervision. However, in their open 12-week study (15), the average minutes engaged (with the videogame therapeutic) dropped by 34% at week 4 and by 50% at week 12 (Table 1). In the Somryst study for chronic insomnia (12), only 60% of subjects completed all 6 core modules of CBT and frequency of subject logins varied from 0-142 times (median of 25). While efficacy was sustained even at 12-month follow up, the decrease in insomnia score was greater in subjects who completed all 6 modules versus those who did not. In the Somryst study for subclinical depression with insomnia (13), attrition rate was 58% at 6 weeks and on average only 3.5 of the 6 modules were completed. Patients completing less than 4 modules had no significant overall benefits versus the control condition and were not different from the control condition at 6 months. In the reSET study for substance use disorder (10), the



drop-out rate was low (12%) – this was likely because subjects were seen twice a week in the clinic, supervised by therapists, paid prizes ranging from "thank you" notes to up to $100 cash for compliance, and on average, earned $277 in prizes over 12 weeks.  In their long-term follow-up (10), when this contingency incentive ended, the superiority of the digital therapeutic over the control condition also ended.

**Closing the Attrition Efficacy Gap**

Mental health conditions, like major depressive disorder, ADHD, and PTSD, require sustained treatment.  Because the field of digital therapeutics is still in its early stages, currently, there is little long-term efficacy data.  If even well-designed, gamified, digital therapeutics have a 50% drop in engagement in 3 months then the outlook for long-term efficacy is grim.  While drop-out rates in clinical trials can be kept low through frequent clinician contact, gamification, feedback, and cash incentives, this is not practical in the real world and hence attrition rates will be far higher.  Finally, if the costs of increasing engagement and compliance equals that of a getting live psychiatric care, then digital therapies would become less attractive as a scalable low-cost solution.

Our scrutiny of published data also reveals several scientific gaps.  First is the lack of standardized definitions of attrition and engagement in the field of digital therapeutics. Second is the lack of standardized reporting requirements by journals.  A single digital therapeutic session can generate a dozen or more different metrics of how a user may interact with the app.  Even widely used clinical trials reporting checklists, such as CONSORT, have not yet required the reporting of all such engagement metrics in digital trials.   This makes it hard to extract such data



from published reports and compare metrics across trials and products.  Third, is the lack of a standardized definition of compliance.  Fourth is the lack of standardized statistical methods, such as mixed models or last observation carried forward, to account for attrition and engagement biases in digital trials.

Fortunately, several constructs are emerging as promising features to increase engagement – both related to external factors of motivation and UX design (24).  Several factors such as ease of use, gamification, ability to personalize app, in-app symptom monitoring, numerical feedback, ability to chart progress, socialization within the app, and integration with clinical services, have been reported to increase engagement (17, 18).  A machine learning analysis of 54,604 adult patients with depression and anxiety identified 5 distinct engagement patterns for digital cognitive behavior therapy over 14-weeks: low engagers [36.5% of sample], late engagers [21.4%], high engagers with rapid disengagement [25.5%], high engagers with moderate decrease [6.0%], and high persistent engagers [10.6%]. Depression improvement rates were lowest for the low engagers (22).  This study suggested machine learning algorithms may be useful to tailor interventions and a human touch for each of these five groups.  Kaiser Permanente found that integrating digital mental health solutions – provided via clinician referral – into their health care delivery system was able to successfully enhance engagement (25).  Fears around privacy and data security for mental health data may be a factor in engagement and attrition for some participants and this should be addressed upfront.  While we do not have all the solutions, encouraging the availability of raw data from clinical trials through trial registries, analyzing long-term real-world data on patient reported outcomes, user experience (engagement and compliance) and product reliability (18) will be important to enhance their utility.



Digital therapeutics for mental health are here to stay. As the pivotal studies demonstrate, they benefit a substantial number of patients. However, the gap between intention and real-world efficacy for digital therapeutics remains large. There is an urgent need to recognize this gap and for stakeholders – regulators, technology developers, clinicians, patients – to come together to close this gap and ensure that this form of treatment is useful to clinical populations.

Acknowledgments/Disclosures:

 This manuscript was not funded by any external source.

AN has no conflicts to disclose. SB has no conflicts to disclose. MMH has received grants from NIMH, NIA, Brain Initiative and Stanley Foundation for other projects. For other projects, PMD has received grants from NIA, DARPA, DOD, ONR, Salix, Avanir, Avid, Cure Alzheimer's Fund, Karen L. Wrenn Trust, Steve Aoki Foundation, and advisory fees from Apollo, Brain Forum, Clearview, Lumos, Neuroglee, Otsuka, Verily, Vitakey, Sermo, Lilly, Nutricia, and Transposon. PMD is a co-inventor on patents for diagnosis or treatment of neuropsychiatric conditions and owns shares in biotechnology companies.

**Table 1. Attrition Rates in Studies of FDA Cleared Digital Therapeutics**

| Reference Number | N | Indication | Study Intervention and Dose | Trial Design | Trial Duration | Engagement | Attrition Rate for Active Intervention Arm |
|---|---|---|---|---|---|---|---|
| 14 | 348 | Pediatric ADHD | RCT of Endeavor Rx vs. control Video game<br><br>25 minutes per day, five days per week, for 4 weeks. | Hybrid | 4 weeks | 83% | 6% |
| 12 | 303 | Chronic Insomnia | SHUTi (Somryst)<br><br>six sequential modules completed within 9 weeks | Remote | 9 weeks, with 12-month follow-up. | 60.3% | 9.2%<br><br>17.5%[1] |
| 15 | 206 | Pediatric ADHD | Open Label study of Endeavor Rx +/- pharmacotherapy<br><br>4 weeks (25 mins/day, 5 days/week), followed by a 4 week pause, and then another 4-week use of therapeutic. | Hybrid | 12 weeks | 68% (pharmacotherapy) 58% (no pharmacotherapy)[2] | 12% (pharmacotherapy)<br><br>12% (no pharmacotherapy) |
| 11 | 170 | Opioid Use Disorder | Therapeutic Education System (reSET-O)<br><br>TES modules 3 times per week for 12 weeks. | In-Clinic | 12 weeks | All sessions were supervised by a live therapist in the clinic. | 20% |
| 10 | 507 | Substance Use Disorder | Therapeutic Education System (reSET)<br><br>4 TES modules per week for 12 weeks. | Hybrid | 12 weeks | 76.3%[3] | 12% |
| 13 | 1149 | Subclinical depression and insomnia | SHUTi (Somryst)<br><br>six sequential modules completed within 6 weeks | Remote | 6 weeks, with 6 month follow up | 58% | 57%<br><br>61%[4] |

[1] Attrition at 12 month follow up. Studies used variable definitions and often did ot break down reasons.
[2] Compliance is defined as the percentage of total possible recommended sessions. Engagement metrics were not reported in a standardized manner
[3] 36.6 modules out of recommended 48 (range 0-72)
[4] Attrition at 6 month follow up.